\documentclass[sigconf]{acmart}

\usepackage{multirow} 
\usepackage[inline]{enumitem} 
\usepackage[ruled]{algorithm2e} 
\usepackage{rotating} 
\setcopyright{none} 
\acmConference[]{}
\acmYear{2021}

\begin{document}

\title{Use of a non-peer reviewed sources in cyber-security scientific research}

\author{Dalibor Gernhardt}
\email{dalibor.gernhardt@gmail.com}
\orcid{0000-0002-5309-3891}
\affiliation{
  \institution{Centre for Defence and Strategic Studies "Janko Bobetko"}
  \institution{Croatian Defense Academy}
  \streetaddress{Ilica 256b}
  \city{Zagreb}
  \country{Croatia}
  \postcode{HR-10000}
}
\author{Stjepan Groš}
\email{stjepan.gros@fer.hr}
\orcid{0000-0001-6619-2859}
\affiliation{%
  \institution{Laboratory for Information Security and Privacy}
  \institution{University of Zagreb Faculty of Electrical Engineering and Computing}
  \streetaddress{Unska 3}
  \city{Zagreb}
  \country{Croatia}
  \postcode{HR-10000}
}

\begin{abstract}
Most publicly available data on cyber incidents comes from private companies and non-academic sources. Common sources of information include various security bulletins, white papers, reports, court cases, and blog posts describing specific events, often from a single point of view, followed by occasional academic sources, usually conference proceedings. The main characteristics of the available data sources are: lack of peer review and unavailability of confidential data. In this paper, we use an indirect approach to identify trusted sources used in scientific work. We analyze how top-rated peer reviewed literature relies on the use of non-peer reviewed sources on cybersecurity incidents. To identify current non-peer reviewed sources on cybersecurity we analyze references in top rated peer reviewed computer security conferences. We also analyze how non-peer reviewed sources are used, to motivate or support research. We examined 808 articles from top conferences in field of computer security. The result of this work are list of the most commonly used non-peer reviewed data sources and information about the context in which this data is used. Since these sources are accepted in top conferences, other researchers can consider them in their future research. To the best of our knowledge, analysis on how non-peer reviewed sources are used in cyber-security scientific research has not been done before. 
\end{abstract}

\ccsdesc[300]{Security and privacy~Social aspects of security and privacy}
\ccsdesc[300]{General and reference~Reference works}
\ccsdesc[300]{General and reference~Empirical studies}

\keywords{non-peer reviewed sources, cyber security, cyber incidents}


\maketitle

\section{Introduction}
\label{sec:introduction}

It is well known that the first information about incidents in cyberspace is usually published in non-peer reviewed sources, such as various news portals, blog posts, or even Twitter feeds, as shown by the recent example of the Solarwinds hack, where users used the hashtag \#sunburst to follow the latest news \cite{Twitter.com}. Only after a certain, possibly long, period of time does the information spread to scientific, i.e. peer-reviewed, sources.

The reason for this state of affairs is that when an incident occurs, various industrial incident managers and handlers are the first on the scene to help the company manage and recover from the incident. In doing so, they will have full access to the compromised systems and all activities taking place at that time. They are also provided with all the necessary resources to assist them in their work, as well as any findings coming from external systems or entities like law enforcement. During this time, some of the information leaks out through the victim themselves. For example, the company may publish some information on company websites for reputation management and control of damage, as seen in example \cite{Ramakrishna2021}. Information also leaks by investigators themselves. They want to show themselves as capable of handling the situation, thus enhancing their reputation among prospective customers. In both cases, the information released is heavily redacted so that nothing considered sensitive is leaked. The more informative leaks come indirectly from people who are familiar with the situation and share less redacted information with journalists or bloggers, or publish it themselves under various nicknames. This is a very dynamic phase, and in addition to these sources, there are many more who simply copy, extract and edit original sources in many other ways and publish them on various portals and social networks. So there is a lot of noise created in a due course. 

Only after the investigation is concluded we get a clearer picture of what happened, though not everything is known even after the investigation concludes and things get again to normal operations. These forensic investigations usually last from 4 to 18 months \cite{Grant2016, Grant2018}. Once the results of the investigations are made public, they are often purged of much of the sensitive but important information.

For these reports to find their way into the academic literature, there is at least several months of additional delay in case of conferences, and for journals the average duration of peer review process could be up to 17 weeks \cite{Huisman2017}. Clearly, the flow of information from industrial sources to academic ones can be slow. 

This situation raises two issues for researchers wishing to study incidents or the threat actors behind them in order to help make defenses better. First, data sources are scattered all over the Internet, as discussed in \cite{Lemay2018}, making it difficult to find them. And second, there is a question about the quality of the sources found. In other words, is it an original source, or was it taken from somewhere? Also, is it all of the information that is out there, or is it only part of it? Was it written by an expert or an expert journalist who knows what the key elements of the incident are and thus gets everything is properly conveyed to the reader, or it was written by someone who doesn't know security so important details were potentially left out?

In this work, we aim to help researchers by identifying relevant and trustworthy non-peer reviewed sources that address computer security issues, such as reports, blogs, and news sites. Researchers can then focus on these sources knowing that the content published in them is more likely to be relevant and can be reliably used to support their research.

Since it is very difficult for a single person to review all the possible sources for its quality, we decided to tackle this problem by studying references in papers published in top security conferences. The idea behind this is that non-peer reviewed sources used by authors that publish in top security conferences are well scrutinized for the quality of its content. In addition to this analysis, we also wanted to know how exactly information from non-peer reviewed sources was used, i.e., whether it was used to support arguments in the main research or it was used only as the motivation for the paper in the introduction section of an article.

Since it is very difficult for a single person to review all the possible sources for its quality, we decided to tackle this problem by studying references in papers published in top security conferences. The idea behind this is that non-peer reviewed sources used by authors that publish in top security conferences are well scrutinized for the quality of its content. In addition to this analysis, we also wanted to know how exactly information from non-peer reviewed sources was used, i.e., whether it was used to support arguments in the main research or it was used only as the motivation for the paper in the introduction section of an article. 

To the best of our knowledge this has not been done before. This work makes the following contributions: first, we have developed an indirect methodology for identification of high-quality non-peer sources, and second, we provide researchers with a list of high-quality, non-peer reviewed sources used in the scientific literature.

This paper is organized as follows. Section \ref{sec:background}. describes what kind of information is publicly available to researchers. Section \ref{sec:methodology} presents our methodology for identification of non-peer reviewed sources in scientific research. Section \ref{sec:datacollection} provides technical insight into our data collection efforts. In Section \ref{sec:results} we present the results of our study, followed by a discussion of the results in Section \ref{sec:discussion}. In Section \ref{sec:recommendations} we provide recommendations for researchers regarding usage of non-peer reviewed sources. Section \ref{sec:conclusion} concludes this paper and gives an overview of the future work.

\section{Background}
\label{sec:background}

The potential sources for information regarding computer security incidents are various reports, newspaper articles, blogs, tweets, threat intelligence (TI), and occasional conference proceedings \cite{Lemay2018}. 

Freely available reports and white papers are often condensed versions of reports available in paid TI \cite{Bouwman2020}. Motives behind releasing free reports can be seen in the marketing and selling companies' security products, services, or TI. The data in free reports and whitepapers is presented in a way that protects the identity of the attack victim and removes any sensitive or confidential information that could be used to link to the victim or the victim's systems. After sanitation, the reports still hold key data, but with limited surrounding context, which makes them difficult to use because not much information is given except for some technical details \cite{Bouwman2020}. It is important to note that these documents have little in common, and they are not standardized or scientifically reviewed in any way. The authors of these documents often do not distinguish between the meaning of the terms report \cite{CambridgeUniversityPressa} and white paper \cite{CambridgeUniversityPress}, and the content of these documents varies depending on the author. Even when researchers have access to paid TI that contain original versions of reports, the use of such documents in scientific research is often not permitted, as discussed in \cite{Bouwman2020}.

Apart from reports, various newspapers and news web portals often cover computer security topics. Some of them are specialized and others cover only the most notable incidents, and thus emphasizing some event. Due to the large number of news websites and portals that often quote each other, it is difficult to assess the true origin of the information. Blogs can be an additional source of information. Blogs can be private, maintained by some well-informed industry insiders, such as \cite{Krebs2021, Schneier2021}, or, as is often the case, maintained by companies. Cybersecurity companies usually have their own dedicated blogs where they publish first-hand information before they are made into specialized reports or white papers.

There are cases where exchange of incident data is mandated at the national level. This data is not publicly available. For example, the European directive on security of network and information systems (NIS directive) aims to share cybersecurity incident data among critical infrastructure companies \cite{THEEUROPEANPARLIAMENTANDTHECOUNCILOFTHEEUROPEANUNION2016, Settanni2017}. The problem with the NIS directive is that only large companies are covered, and even then, only significant incident data is shared, where definition of a significant incident depends from case to case on many factors \cite{EuropeanUnionAgencyforCybersecurityENISA2021}. 

Other data sources can be seen in the form of free Open sources of Threat Intelligence (OTI) such as \cite{ATTbuisiness}. The general characteristic of OTIs is that they are highly technical and rely on references to other data sources, sources such as those we seek to identify in this work. As such, OTIs are not data sources in their own right; from the perspective of this work, they can be seen as repositories of pointers to other data sources.

If this is the case, one may wonder where do peer-reviewed articles get their input information from. It is known that there are numerous reliable and fast sources dealing with this type of news: both personal and security companies blogs, various reports, etc., but since they are dispersed there is little knowledge about the quality of the data provided. In the following sections, we aim to identify publicly available and non-peer reviewed data sources exists and assess how this data is used in scientific work.

\section{Methodology}
\label{sec:methodology}

Since we do not have the resources to analyze all available data sources: reports, various news and blog sites, to find out which of them publish high quality original content, our goal was to derive a methodology for identifying those sources in a different manner. We decided to take an indirect approach: We selected a set of trustworthy data sources and analyzed which sources are used in these documents. 

Typically, as the prestige of a journal or conference increases, so does the quality control of its content. This implies that these papers must have used high quality references as sources. Top rated journals and conferences have stricter rules regarding the quality of sources compared to lower rated ones. In other words: We leave it to the reviewers to filter out trustworthy sources of information about computer security incidents.

Since the peer review time of journals tends to be longer compared to conferences, we have chosen to analyze top computer security conferences. The second argument for using conferences rather than journals is the assumption that the nature of these sources is changing. From our experience, we have found that non-peer reviewed sources change over time, one day they are active with lots of new content and the next day they are gone. This change over time means that there is no guarantee that a source used in the past will still provide quality content today. Our goal is to find out which are the best sources at the present time. Our methodology can be broken down into the following steps:

\begin{enumerate}
    \item[Step 1:] \textit{Identify top peer-reviewed conferences.}
        The first step in the process is to identify the relevant rankings that show the top conferences in a field of computer security.

    \item[Step 2:] \textit{Identify commonly used non-peer reviewed sources.}
        Our initial assumption is that most, if not all, non-peer reviewed sources are referenced by a URL. This information can be used to isolate all URLs in reference sections of conference proceedings. We are aware that this method will not cover all cases of non-peer reviewed sources, as some of them are not available online. To mitigate this, we will perform a manual search of the dataset using a set of keywords that might contain such sources.
        
    \item[Step 3:] \textit{Reference usage analysis.}
    
        After identifying the most commonly used non-peer reviewed sources, we analyzed how these references were used in the articles. The goal was to determine if the references were used to support motivation or if they contained information that was relevant to the research itself. For the purpose of this research, we divided the articles into the following distinguishable sections:
        
        \begin{enumerate}
            \item introduction and related work, 
            \item background, 
            \item research, 
            \item discussion.
        \end{enumerate}

\end{enumerate}

After we have done all the three steps, the result is a list of non-peer reviewed sources of the highest quality. These sources are trusted enough to be used as sources in papers at top conferences, which means they contain trustworthy data. An additional contribution is the analysis that shows how these sources were used. Sources used in the main research indicate that the information provided is relevant to the research itself, and data used in other parts may indicate that this information was a trigger for some research.

\section{Data collection}
\label{sec:datacollection}

The first step in data collection is to determine the ranking of relevant conferences. Once the relevant conferences are identified, we need to decide how many conferences with the highest ranking should be analyzed and set a time frame for the analysis, as described in Section \ref{subsec:topconference}. The next step is to identify all non-peer reviewed data sources used in the reference sections of the conference papers (Section \ref{subsec:identification}). Section \ref{subsec:reference} describes the technical difficulties encountered in identifying the non-peer reviewed sources used in the papers. We conclude this section by presenting observations on trends in the computer industry in Section \ref{subsec:observations}: various security companies acquisitions and name changes, and the tendency of security companies to publish data using multiple brands. These events have influenced our analysis and can be expected in the future.

\subsection{Top computer security conferences}
\label{subsec:topconference}

To identify the most relevant computer security conferences, we used Microsoft Academic's saliency ranking \cite{Microsoft2021} . We decided to analyze the top 9 computer security conferences in the period of one year - 2020, as shown in Table \ref{table1}. Originally, we intended to analyze the top 10 conferences, but we were not able to obtain all the proceedings of the tenth conference on the list, so we removed it from our research.

\begin{table}
\centering
\caption{Top conferences in computer security in 2020 according to Microsoft Academic Research \cite{Microsoft2021}}

\label{table1}
\begin{tabular}{|c|p{0.6\linewidth}|} 
\hline

\textbf{Rank} & \textbf{Conference} \\ 
\hline
\hline
1 & IEEE Symposium on Security and Privacy (S\&P) 2020 \\ 
\hline
2 & CCS’20~ ACM SIGSAC Conference on Computer and Communications Security \\ 
\hline
3 & USENIX Security Symposium '20 \\ 
\hline
4 & SIGCOMM '20: Annual conference of the ACM -- Special Interest Group on Data Communication on the applications, technologies, architectures, and protocols for computer communication \\ 
\hline
5 & 40th Annual International Cryptology Conference -- CRYPTO 2020 \\ 
\hline
6 & Network and Distributed System Security Symposium 2020 (NDSS) \\ 
\hline
7 & Financial Cryptography and Data Security (FC) 2020 \\ 
\hline
8 & Annual Computer Security Applications Conference (ACSAC) 2020 \\ 
\hline
9 & IEEE International Conference on Computer Communications (ICC) 2020 \\
\hline
\end{tabular}
\end{table}

Once we identified the conferences and the desired time period for analysis, we created a local dataset containing 570 PDF files with a total of 808 articles. Here we can see that some conferences offer the option for downloading individual papers, while others tend to publish conference proceedings as books containing hundreds of papers. This inconsistency regarding how conference proceedings are made available complicates automation in future steps of our research, as it is difficult to automate the analysis of differently structured data.

\subsection{Identification of relevant non-peer reviewed data sources.}
\label{subsec:identification}

Our initial assumption was that most, if not all, non-peer reviewed sources should be cited with associated URLs. With this assumption in mind, we have searched our entire dataset for the strings "http://" and "https://" and extracted the surrounding text to obtain the full URL and some context. 

While preparing for this phase, we evaluated several approaches on how to analyze this data. One idea was to search the URL strings for keywords. This approach was discarded because we could not predict all possible keywords and, finally, there is a possibility that the content of the URL string does not correlate with the content of the web page. Finally, we decided to perform an analysis of how frequently each URL appears. Once frequent URLs are identified, then we will proceed to evaluation of webpages.  During the analysis, we did not rely on existing lists of web page classifications such as those used to filter internet content, e.g., \cite{VirtualGraffitiInc2021} , and opted for a manual analysis of the content referenced by each URL that appears above a certain threshold. In our initial search of the dataset, we identified 12560  URLs. To identify commonly used sources, we have grouped the URLs by their associated domain names, e.g., we found 8 different URLs pointing to the domain "krebsonsecurity.com". This process of grouping URLs to domain names resulted in the identification of 237 domain names that were cited more than five times. 

Closer examination of the frequently occurring domain names revealed that most of the cited URLs were actually links to different Digital Object Identifiers (DOIs), various article, or source code repositories, rather than non-peer reviewed sources. To identify relevant non-peer reviewed sources, we excluded domains associated with the following topics: DOIs, articles and proceedings repositories, source code repositories, vulnerability databases, exploit databases, news about various products and corresponding updates, Wikipedia articles, technical specifications, manuals, cryptocurrency news, various statistics and metrics, protocol documentation, processor documentation, anonymity browsing, bug databases, as these types of sources are not relevant to our research. 

Through this search, we only identified sources that are cited using URLs. If some source is cited without a URL, it would not be identified automatically. For example, non-peer reviewed conferences (such as Black Hat) often provide first-hand information on hacking and cybersecurity. From our experience, we have found that authors of articles cite these sources in inconsistent ways, with some authors citing with the URL and others citing with the name of the conference and authors. Since our primary search is focused on URLs, these types of results would not show up unless they were cited with URL. To include these types of results, we used several lists of hacking conferences and searched our dataset for mentions of conference names in reference sections and included them in our list. In addition to conferences, we manually searched the dataset for keywords such as "SANS Institute", reputable sources that could be used as sources but are not cited using URL.

To narrow down the list to relevant sources, we decided to analyze only sources that were cited five or more times. If some source has a frequency less than five, then we consider it rarely used and dismiss it. This process resulted in the elimination of about 5 500 rarely cited domains or keywords. After examining the content referenced by the URLs, we obtained a list of domains related to various computer security topics: specialized news sites, blogs, industrial sources publishing various reports and white papers.

\subsection{Identifying the reference context}
\label{subsec:reference}

After reducing the list of sources to the most relevant domain names, we searched the dataset again, this time using each relevant keyword identified in a previous step. We evaluated where in the paper a reference to a particular source was used. This was mostly manual and time-consuming step for several reasons:

\begin{description}[style=nextline]
    \item[i. Number of queries] \hfill After the process of grouping URL to frequent domain names and adding additional keywords for the search as described in the previous section, we performed 150 queries over 808 papers.
    \item[ii. Inconsistency of article structure] \hfill Each conference uses a different template, each author has a different number of sections in the paper. Also, as mentioned in IV.A, some conferences allow access to individual papers, and some publish papers in the form of a book.
    \item[iii. Different citation rules] \hfill Some citation rules state that sources are sorted alphabetically, while others state that sources are sorted in the order in which they appear in the paper. When authors cite multiple sources at once, the references are often grouped, e.g., "[1-5]", with the reference number hidden in range of values.
\end{description}

In all the cases, our process was as follows: first, we searched the paper for the keywords, if the certain keyword was found, then we identified the reference number in the reference section of the paper, e.g., " [19]", third, we evaluated where in the paper a reference to a particular source was used.

Distinction of sections in papers is difficult to standardize objectively and is therefore susceptible to subjective interpretation and questioning. The primary goal of this paper was to determine what currently used non-peer reviewed sources exist, and the secondary goal was to evaluate the context in which they were used. The authors made every effort to be as objective as possible and the results provide at least some indication of how these sources are used.

\subsection{Observations}
\label{subsec:observations}

During our research, we came across several cases where cited source originated from a security company that was acquired by another company at some point, resulting in both brand names being cited depending on when the author accessed the content or when post was originally published. For example, since Symantec was acquired by Broadcom in 2019 \cite{Riley2019}, publications are published under the different brand - Broadcom, resulting in citations of both names. To further complicate the analysis, Broadcom publishes different types of documents from different domain names, such as: blogs, white papers, reports. Similarly, Palo Alto Networks acquired PureSec in 2019 \cite{PaloAltoNetworks2019}, AT\&T acquired AlienVault \cite{ATTBuisiness2018}. 

Another observation is that large security companies post documents under different brand names. Sophos frequently posts news under the brand name NakedSecurity, Eset posts under the brand name WeLiveSecurity, Kaspersky uses the brand name SecureList, Cisco uses both  Cisco and Talos names, and Microsoft uses Technet, MSDN, and Azure names. 

Because the citations from these sources are spread across multiple brands with the same origin, in some cases with only a few citations from individual sources, we grouped the results from the sources and presented them as one data source, grouping both the names and the types of reports. The only exception to this rule were sources originating from Google. Both "security.googleblog.com" and "googleprojectzero.blogspot.com" were cited more than 5 times, so we decided to treat them as separate sources.

\section{Results}
\label{sec:results}

The results are presented as follows: In Section \ref{subsec:distribution} we present how often different conferences rely on non-peer reviewed sources. Full results of our study showing which sources are most frequently used are presented in Section \ref{subsec:identify}, and in Section \ref{subsec:analysis} we present our findings on usage of non-peer reviewed sources regarding to a paper context.

\subsection{Distribution of non-peer reviewed sources across conferences.}
\label{subsec:distribution}

After analysis of top nine conferences in the field of computer security during 2020 (Table \ref{table1}), we have noticed uneven distribution in usage of non-peer reviewed sources from conference to conference. Possible reason for this is the fact that all conferences are in the field of computer security, but focus on different topics. Table \ref{table2}. summarizes the use of non-peer reviewed sources at each conference. It is noticeable that the conferences SIGCOMM, CRYPTO, FC, and ICC have a total of 14 references to non-peer reviewed review sources that meet our criteria. 

\begin{table}
\centering
\caption{Distribution of non-peer reviewed citations among top nine conferences}
\begin{tabular}{|c|c|c|} 
\hline
\textbf{Conference} & \begin{tabular}[c]{@{}c@{}}\textbf{ Number of }\\\textbf{non-peer }\\\textbf{reviewed }\\\textbf{citations }\end{tabular} & \begin{tabular}[c]{@{}c@{}}\textbf{ Percentage of total }\\\textbf{number }\end{tabular}  \\ 
\hline
\hline
SP 2020 & 103 & 23 \% \\ 
\hline
CCS ’20 & 96 & 21 \% \\ 
\hline
Security '20 & 101 & 23 \% \\ 
\hline
SIGCOMM '20 & 10 & 2 \% \\ 
\hline
CRYPTO 2020 & 1 & 0 \% \\ 
\hline
NDSS 2020 & 77 & 17 \% \\ 
\hline
FC 2020 & 1 & 0 \% \\ 
\hline
ACSAC 2020 & 57 & 13 \% \\ 
\hline
ICC 2020 & 2 & 0 \%  \\ 
\hline
\textbf{Total:}     & \textbf{448} & ~  \\
\hline
\end{tabular}
\label{table2}
\end{table}

The small sample size of the four conferences is not sufficient to draw firm conclusions. For this reason, we excluded these four conferences from further analysis. This left us with data from five conferences that are relevant to the field and that use non-peer reviewed sources: S\&P, CCS, Security, NDSS, and ACSAC.

\subsection{Identifying non-peer reviewed data sources}
\label{subsec:identify}

Through our analysis of references used in the top 2020 cybersecurity conferences, we identified 37 data sources that were cited more than 5 times, for a total of 434 times in 540 articles from five conferences, as shown in Table \ref{table3}.

\begin{table}
\centering
\caption{Specialized non-peer reviewed sources used in 5 analyzed conferences held in 2020}
\begin{tabular}{|c|c|c|c|} 
\hline
\textbf{No} & \textbf{Source }          & \textbf{Type}      & \begin{sideways}\textbf{No. of citations }\end{sideways} \\ 
\hline
\hline
1. & Blackhat series                & Conference         & 54 \\ 
\hline
2. & googlepro...blogspot.com* & Sec. comp. blog & 27 \\ 
\hline
3. & zdnet.com                      & Specialized news   & 25 \\ 
\hline
4. & wired.com                      & Specialized news   & 24 \\ 
\hline
5. & Broadcom (Symantec)            & Security company      & 20 \\ 
\hline
6. & arstechnica.com                & Specialized news   & 19 \\ 
\hline
7. & techcrunch.com                 & Specialized news   & 18 \\ 
\hline
8. & theguardian.com                & General news       & 14 \\ 
\hline
9. & theverge.com                   & General news       & 13 \\ 
\hline
10. & forbes.com                     & General news       & 13 \\ 
\hline
11. & security.googleblog.com        & Sec. comp. blog & 13 \\ 
\hline
12. & grsecurity.net                 & Security company      & 12 \\ 
\hline
13. & Kaspersky (Securelist)         & Security company      & 12 \\ 
\hline
14. & Fireeye                        & Security company      & 11 \\ 
\hline
15. & bbc.com                        & General news       & 10 \\ 
\hline
16. & BleepingComputer.com           & Specialized news   & 10 \\ 
\hline
17. & bits-please.blogspot.com       & Private blog       & 9 \\ 
\hline
18. & Microsoft (Azure, MSDN)        & Security company      & 9 \\ 
\hline
19. & reuters.com                    & General news       & 8 \\ 
\hline
20. & SANS Institute                 & Institute          & 8 \\ 
\hline
21. & KrebsOnSecurity.com            & Private news       & 8 \\ 
\hline
22. & Palo Alto Networks             & Security company      & 8 \\ 
\hline
23. & Sophos (Naked security)        & Security company      & 8 \\ 
\hline
24. & Cisco (Talos)                  & Sec. comp. blog & 8 \\ 
\hline
25. & DEF CON                        & Conference         & 7 \\ 
\hline
26. & washingtonpost.com             & General news       & 7 \\ 
\hline
27. & theregister.co.uk              & General news       & 7 \\ 
\hline
28. & citizenlab.ca                  & Laboratory         & 6 \\ 
\hline
29. & thehackernews.com              & Specialized news   & 6 \\ 
\hline
30. & dyn.com/blog                   & Sec. comp. blog & 6 \\ 
\hline
31. & telegraph.co.uk                & General news       & 5 \\ 
\hline
32. & bloomberg.com                  & General news       & 5 \\ 
\hline
33. & McAfee report                  & Security company      & 5  \\ 
\hline
34. & blog.cloudﬂare.com             & Sec. comp. blog & 5 \\ 
\hline
35. & TrendMicro Blog                & Sec. comp. blog & 5 \\ 
\hline
36. & signal.org/blog/               & Sec. comp. blog & 5 \\ 
\hline
37. & blog.apnic.net**                & Sec. comp. blog & 4  \\ 
\hline
\multicolumn{3}{|c|}{\textbf{Total count }} & \textbf{434} \\ 
\hline
\multicolumn{4}{l}{\begin{tabular}[c]{@{}l@{}}*google.projectzero.com \\ **blog.apnic.net falls under source 5 because it was once cited \\in SIGCOMM '20 which was not analyzed regarding \\position of citation in article\end{tabular}}   
\end{tabular}
\label{table3}
\end{table}

As described in Section IV.B, we identified data sources using two methods: analysis of URL string analysis and searching references against a list of possible sources which are not cited by URL. Our study shows that out of 37 frequent sources, 34 were cited by URL (92\%), with only three sources that were not cited by URL: Blackhat and DEFCON conferences, and the SANS Institute. If we focus on the quantity of citations, the results are as follows: Out of 434 citations, 365 were cited by URLs (84\%). 

If we focus on the type of sources rather than the number of citations when interpreting the results in Table \ref{table3}. the results are as follows: The most frequently cited sources are unspecialized but highly reputable news websites (24\%), followed by security companies publications (21\%) and security companies blog posts (21\%). In fourth place are specialized news sites (16\%), with all other types: conferences, laboratories, private blogs and news, institutes falling into the remaining 18\%.

If we focus instead on the number of references originating from different types of sources, the results are as follows: Specialized news websites are the most cited with 102 citations or 24\%, followed by security companies publications with 85 (20\%) citations. Non-specialized news websites are third with 82 (19\%) citations, security company blogs are fourth with 73 (17\%) citations, and hacking conferences are fifth with 61 citations (14\%), with all other sources falling within the remaining 6\%.

\subsection{Analysis of reference context}
\label{subsec:analysis}

The final goal of this work was to find out in which contexts authors used non-peer reviewed sources and whether the use of these sources was consistent from conference to conference. We wanted to know if these sources were used only to motivate the work, to show that a particular behavior was found in practice, or if these data were used in the main body of the research. As described in Section \ref{sec:methodology}, we manually analyzed 434 citations with the goal of determining how these sources were used.

The results of this analysis are presented in Table \ref{table4}. showing which non-peer reviewed sources are used, their type, how they are used in papers regarding context, both at the level of individual conference observed conference and on average. The results shown in \ref{table5}. show distribution of non-peer reviewed references in all conferences: Non-peer reviewed sources were most frequently cited in the introductory sections of articles or in the review of related work at 61\% of the time. In the development section, these sources are used in 17\% of citations and in the main body of the research in 16\%. As shown in Table \ref{table5} the use of references in discussions is not consistent between the observed conferences.

\begin{sidewaystable*}
\centering
\small 
\caption{Summarized results of relevant non-peer reviewed sources in computer security conferences of 2020}
\begin{tabular}{|c|l|l|c|c|c|c|c|c|c|c|c|c|c|c|c|c|c|c|c|c|c|c|c|c|c|c|c|}
\hline
\multirow{2}{*}{~} & \multirow{2}{*}{Cites Source} & \multirow{2}{*}{Type} & \multicolumn{4}{c}{S\&P 2020} \vline & \multicolumn{4}{c}{CCS '20} \vline & \multicolumn{4}{c}{Security '20} \vline & \multicolumn{4}{c}{NDSS 2020} \vline & \multicolumn{4}{c}{ACSAC 2020} \vline & \multirow{2}{*}{$\sum A$} & \multirow{2}{*}{$\sum B$} & \multirow{2}{*}{$\sum C$} & \multirow{2}{*}{$\sum D$} & \multirow{2}{*}{$\sum$} \\ 
\cline{4-23}
  &   &   & A  & B  & C  & D & A  & B  & C  & D & A & B & C  & D & A  & B  & C  & D & A & B & C & D &   &   &   &   &   \\ 
\hline
\hline
1. & Blackhat series & Conference & 5  & ~  & 4  & ~ & 5  & 3  & 4  & ~ & 13 & 3  & 3  & ~ & 7  & 2  & 1  & ~ & 3  & 1 & ~ & ~  & 33 & 9 & 12 & 0 & 54 \\ 
\hline
2. & googleprojectzero & Sec. comp. blog                                                       & 2  & 1  & 1  & 1                                             & 5  & 1  & 1  & ~                                           & 9  & 1  & ~  & 3                                                 & 2  & ~  & ~  & ~                                             & ~  & ~ & ~ & ~                                                & 18                                 & 3                                  & 2                                  & 4                                  & 27 \\ 
\hline
3. & zdnet.com                                                  & Secialized news                                                          & 1  & ~  & ~  & 2                                             & 5  & 1  & 1  & 1                                           & 4  & ~  & ~  & ~                                                 & 1  & ~  & 2  & ~                                             & 4  & 1 & 2 & ~                                                & 15                                 & 2                                  & 5                                  & 3                                  & 25 \\ 
\hline
4. & wired.com                                                  & Secialized news                                                          & 7  & ~  & 1  & ~                                             & 5  & 1  & 1  & ~                                           & 4  & ~  & ~  & ~                                                 & 2  & ~  & ~  & ~                                             & 3  & ~ & ~ & ~                                                & 21                                 & 1                                  & 2                                  & 0                                  & 24  \\ 
\hline
5. & Broadcom (Symantec)                                        & Security comp.                                                            & 5  & ~  & ~  & ~                                             & 2  & ~  & ~  & ~                                           & 4  & ~  & ~  & ~                                                 & 3  & 1  & ~  & ~                                             & 4  & 1 & ~ & ~                                                & 18                                 & 2                                  & 0                                  & 0                                  & 20 \\ 
\hline
6. & arstechnica.com                                            & Secialized news                                                          & 5  & 1  & 1  & ~                                             & 3  & 1  & 1  & ~                                           & ~  & ~  & 2  & ~                                                 & 3  & 1  & ~  & 1                                             & ~  & ~ & ~ & ~                                                & 11                                 & 3                                  & 4                                  & 1                                  & 19 \\ 
\hline
7. & techcrunch.com                                             & Secialized news                                                          & 3  & 1  & 1  & ~                                             & 1  & 1  & 1  & ~                                           & 3  & 1  & 1  & 1                                                 & 1  & ~  & ~  & ~                                             & ~  & ~ & 3 & ~                                                & 8                                  & 3                                  & 6                                  & 1                                  & 18  \\ 
\hline
8. & theguardian.com                                            & General news                                                             & ~  & ~  & 1  & ~                                             & 1  & 1  & 2  & ~                                           & 3  & ~  & ~  & 1                                                 & 2  & ~  & ~  & 1                                             & 1  & 1 & ~ & ~                                                & 7                                  & 2                                  & 3                                  & 2                                  & 14                                \\ 
\hline
9. & forbes.com                                                 & General news                                                             & ~  & 2  & 1  & ~                                             & 2  & 1  & ~  & 3                                           & ~  & 1  & 1  & ~                                                 & ~  & 1  & ~  & ~                                             & 1  & ~ & ~ & ~                                                & 3                                  & 5                                  & 2                                  & 3                                  & 13                                \\ 
\hline
10. & security.googleblog.com                                    & Sec. comp. blog                                                       & 1  & ~  & 1  & 1                                             & ~  & ~  & 2  & ~                                           & 3  & 1  & ~  & ~                                                 & 3  & ~  & ~  & 1                                             & ~  & ~ & ~ & ~                                                & 7                                  & 1                                  & 3                                  & 2                                  & 13                                \\ 
\hline
11. & theverge.com                                               & General news                                                             & 1  & 2  & ~  & ~                                             & 1  & ~  & 2  & ~                                           & 2  & ~  & ~  & 1                                                 & ~  & 1  & ~  & ~                                             & 2  & ~ & 1 & ~                                                & 6                                  & 3                                  & 3                                  & 1                                  & 13                                \\ 
\hline
12. & grsecurity.net                                             & Security comp.                                                            & 2  & 3  & ~  & ~                                             & 1  & 1  & ~  & ~                                           & 4  & ~  & ~  & ~                                                 & ~  & 1  & ~  & ~                                             & ~  & ~ & ~ & ~                                                & 7                                  & 5                                  & 0                                  & 0                                  & 12                                \\ 
\hline
13. & Kaspersky (Securelist)                                     & Security comp.                                                            & ~  & ~  & ~  & ~                                             & 5  & ~  & ~  & ~                                           & 2  & ~  & ~  & ~                                                 & 2  & 1  & ~  & ~                                             & 1  & 1 & ~ & ~                                                & 10                                 & 2                                  & 0                                  & 0                                  & 12                                \\ 
\hline
14. & Fireeye                                                    & Security comp.                                                            & 1  & 1  & ~  & ~                                             & ~  & ~  & ~  & ~                                           & 2  & 3  & ~  & ~                                                 & 1  & ~  & ~  & 1                                             & 2  & ~ & ~ & ~                                                & 6                                  & 4                                  & 0                                  & 1                                  & 11                                \\ 
\hline
15. & BleepingComputer.com                                       & Secialized news                                                          & 1  & ~  & 2  & ~                                             & ~  & ~  & ~  & ~                                           & 2  & ~  & ~  & ~                                                 & ~  & ~  & 1  & 1                                             & 1  & ~ & 2 & ~                                                & 4                                  & 0                                  & 5                                  & 1                                  & 10                                \\ 
\hline
16. & bbc.com                                                    & General news                                                             & 3  & ~  & 1  & 1                                             & ~  & ~  & ~  & ~                                           & ~  & ~  & 1  & ~                                                 & ~  & 1  & 1  & ~                                             & 2  & ~ & ~ & ~                                                & 5                                  & 1                                  & 3                                  & 1                                  & 10                                \\ 
\hline
17. & Microsoft (Azure, MSDN)                                    & Security comp.                                                            & 2  & 1  & 1  & 2                                             & 1  & ~  & ~  & ~                                           & ~  & ~  & ~  & ~                                                 & ~  & 1  & ~  & ~                                             & ~  & 1 & ~ & ~                                                & 3                                  & 3                                  & 1                                  & 2                                  & 9                                 \\ 
\hline
18. & bits-please.blogspot.com                                   & Private blog                                                             & 7  & 2  & ~  & ~                                             & ~  & ~  & ~  & ~                                           & ~  & ~  & ~  & ~                                                 & ~  & ~  & ~  & ~                                             & ~  & ~ & ~ & ~                                                & 7                                  & 2                                  & 0                                  & 0                                  & 9                                 \\ 
\hline
19. & Palo Alto Networks                                         & Security comp.                                                            & ~  & ~  & ~  & 1                                             & 1  & 2  & ~  & ~                                           & ~  & 1  & ~  & ~                                                 & ~  & ~  & ~  & 1                                             & 1  & 1 & ~ & ~                                                & 2                                  & 4                                  & 0                                  & 2                                  & 8                                 \\ 
\hline
20. & SANS Institute                                             & Institute                                                                & ~  & ~  & ~  & ~                                             & ~  & ~  & ~  & ~                                           & ~  & ~  & 2  & ~                                                 & 1  & 2  & ~  & 1                                             & 2  & ~ & ~ & ~                                                & 3                                  & 2                                  & 2                                  & 1                                  & 8                                 \\ 
\hline
21. & reuters.com                                                & General news                                                             & 1  & ~  & ~  & ~                                             & 2  & 1  & 1  & ~                                           & ~  & ~  & 1  & ~                                                 & ~  & ~  & 1  & ~                                             & 1  & ~ & ~ & ~                                                & 4                                  & 1                                  & 3                                  & 0                                  & 8                                 \\ 
\hline
22. & Cisco (Talos)                                              & Sec. comp. blog                                                       & ~  & 1  & 1  & ~                                             & 1  & ~  & ~  & ~                                           & 1  & 2  & ~  & ~                                                 & ~  & ~  & ~  & ~                                             & 2  & ~ & ~ & ~                                                & 4                                  & 3                                  & 1                                  & 0                                  & 8                                 \\ 
\hline
23. & Sophos (Naked security)                                    & Security comp.                                                            & 2  & ~  & ~  & ~                                             & 2  & ~  & 1  & ~                                           & 1  & ~  & ~  & ~                                                 & 1  & ~  & ~  & ~                                             & 1  & ~ & ~ & ~                                                & 7                                  & 0                                  & 1                                  & 0                                  & 8                                 \\ 
\hline
24. & KrebsonSecurity.com                                        & Private news                                                             & 3  & ~  & ~  & ~                                             & 1  & ~  & ~  & ~                                           & 1  & ~  & ~  & ~                                                 & 2  & ~  & ~  & ~                                             & ~  & 1 & ~ & ~                                                & 7                                  & 1                                  & 0                                  & 0                                  & 8                                 \\ 
\hline
25. & theregister.co.uk                                          & General news                                                             & 1  & ~  & ~  & 1                                             & ~  & ~  & ~  & ~                                           & 3  & ~  & ~  & ~                                                 & 1  & ~  & ~  & ~                                             & 1  & ~ & ~ & ~                                                & 6                                  & 0                                  & 0                                  & 1                                  & 7                                 \\ 
\hline
26. & DEF CON                                                    & Conference                                                               & ~  & ~  & 1  & ~                                             & ~  & ~  & 1  & ~                                           & ~  & ~  & ~  & ~                                                 & 2  & 2  & 1  & ~                                             & ~  & ~ & ~ & ~                                                & 2                                  & 2                                  & 3                                  & 0                                  & 7                                 \\ 
\hline
27. & washingtonpost.com                                         & General news                                                             & ~  & ~  & 1  & ~                                             & 2  & ~  & ~  & ~                                           & ~  & 1  & ~  & ~                                                 & ~  & 1  & ~  & ~                                             & 2  & ~ & ~ & ~                                                & 4                                  & 2                                  & 1                                  & 0                                  & 7                                 \\ 
\hline
28. & citizenlab.ca                                              & Laboratory                                                               & 1  & ~  & ~  & ~                                             & 1  & ~  & ~  & ~                                           & ~  & ~  & ~  & ~                                                 & 1  & ~  & 1  & 1                                             & 1  & ~ & ~ & ~                                                & 4                                  & 0                                  & 1                                  & 1                                  & 6                                 \\ 
\hline
29. & dyn.com/blog                                               & Sec. comp. blog                                                       & 1  & ~  & ~  & ~                                             & 1  & ~  & ~  & ~                                           & ~  & ~  & ~  & ~                                                 & 2  & 1  & 1  & ~                                             & ~  & ~ & ~ & ~                                                & 4                                  & 1                                  & 1                                  & 0                                  & 6                                 \\ 
\hline
30. & thehackernews.com                                          & Secialized news                                                          & 1  & ~  & ~  & ~                                             & ~  & 1  & ~  & ~                                           & ~  & ~  & ~  & ~                                                 & ~  & 1  & ~  & ~                                             & 3  & ~ & ~ & ~                                                & 4                                  & 2                                  & 0                                  & 0                                  & 6                                 \\ 
\hline
31. & signal.org/blog/                                           & Sec. comp. blog                                                       & ~  & 1  & ~  & ~                                             & 2  & ~  & ~  & ~                                           & ~  & ~  & 1  & ~                                                 & ~  & ~  & 1  & ~                                             & ~  & ~ & ~ & ~                                                & 2                                  & 1                                  & 2                                  & 0                                  & 5                                 \\ 
\hline
32. & blog.cloudﬂare.com                                         & Sec. comp. blog                                                       & ~  & ~  & ~  & ~                                             & ~  & ~  & ~  & ~                                           & 2  & ~  & 2  & ~                                                 & 1  & ~  & ~  & ~                                             & ~  & ~ & ~ & ~                                                & 3                                  & 0                                  & 2                                  & 0                                  & 5                                 \\ 
\hline
33. & TrendMicro Blog                                            & Sec. comp. blog                                                       & 1  & 1  & ~  & ~                                             & 2  & ~  & ~  & ~                                           & ~  & ~  & ~  & ~                                                 & ~  & ~  & ~  & ~                                             & ~  & 1 & ~ & ~                                                & 3                                  & 2                                  & 0                                  & 0                                  & 5                                 \\ 
\hline
34. & telegraph.co.uk                                            & General news                                                             & ~  & ~  & ~  & ~                                             & ~  & 1  & ~  & ~                                           & 1  & ~  & ~  & ~                                                 & 2  & ~  & ~  & ~                                             & 1  & ~ & ~ & ~                                                & 4                                  & 1                                  & 0                                  & 0                                  & 5                                 \\ 
\hline
35. & McAfee report                                              & Security comp.                                                            & 1  & ~  & ~  & ~                                             & ~  & 1  & ~  & ~                                           & 3  & ~  & ~  & ~                                                 & ~  & ~  & ~  & ~                                             & ~  & ~ & ~ & ~                                                & 4                                  & 1                                  & 0                                  & 0                                  & 5                                 \\ 
\hline
36. & bloomberg.com                                              & General news                                                             & 1  & ~  & ~  & ~                                             & 2  & ~  & ~  & ~                                           & ~  & ~  & ~  & ~                                                 & 1  & ~  & ~  & ~                                             & 1  & ~ & ~ & ~                                                & 5                                  & 0                                  & 0                                  & 0                                  & 5                                 \\ 
\hline
37. & blog.apnic.net* & Sec. comp. blog & ~  & ~  & ~  & ~ & 2  & ~  & 1  & ~ & ~  & ~  & ~  & ~ & 1  & ~  & ~  & ~ & ~  & ~ & ~ & ~ & 3 & 0 & 1 & 0 & 4 \\ 
\hline
\hline
\multicolumn{3}{r}{TOTAL} & 59  & 17  & 18  & 9 & 56  & 17  & 19  & 4 & 67  & 14  & 14  & 6 & 42  & 17  & 10  & 8 & 40  & 9 & 8 & 0 & 264 & 74 & 69 & 27 & 434 \\
\hline
\multicolumn{27}{l}{}  \\
\multicolumn{27}{l}{*A - Introduction or Related Work, B - Development, C - Main reaseach, D - Dicussion}  \\
\multicolumn{27}{l}{Numbers show how many times source was cited in certain part of article}  \\
\end{tabular}
\label{table4}
\end{sidewaystable*}

\begin{table}
\centering
\caption{Distribution of reference usage regarding a position within articles}
\begin{tabular}{|c|c|c|c|c|c|} 
\hline
\begin{sideways}\textbf{Conference}\end{sideways} & \begin{sideways}\textbf{Introduction or Related work}\end{sideways} & \begin{sideways}\textbf{Development}\end{sideways} & \begin{sideways}\textbf{Main research}\end{sideways} & \begin{sideways}\textbf{Discussion}\end{sideways} & \begin{sideways}\textbf{References}\end{sideways}  \\ 
\hline
\hline
\textbf{ S\&P 2020 }                              & 57\%                                                                  & 17\%                                                 & 17\%                                                   & 9\%                                                 & 103                                                  \\ 
\hline
\textbf{ CCS ’20 }                                & 58\%                                                                  & 18\%                                                 & 20\%                                                   & 4\%                                                 & 96                                                   \\ 
\hline
\textbf{ Security '20 }                           & 66\%                                                                  & 14\%                                                 & 14\%                                                   & 6\%                                                 & 101                                                  \\ 
\hline
\textbf{ NDSS 2020 }                              & 55\%                                                                  & 22\%                                                 & 13\%                                                   & 10\%                                                & 77                                                   \\ 
\hline
\textbf{ ACSAC 2020 }                             & 70\%                                                                  & 16\%                                                 & 14\%                                                   & 0\%                                                 & 57                                                   \\ 
\hline
\textbf{Average}                                  & \textbf{ 61\% }                                                       & \textbf{ 17\% }                                      & \textbf{ 16\% }                                        & \textbf{ 6\% }                                      & \textbf{ 86.8 }                                      \\ 
\hline
\textbf{ Standard dev. }                          & 6.6                                                                   & 3.1                                                  & 2.9                                                    & 4.1                                                 & 19.6                                                 \\
\hline
\end{tabular}
\label{table5}
\end{table}

\section{Discussion}
\label{sec:discussion}

This section is organized as follows: In Section \ref{subsec:independent} we analyze independent researchers used as sources in papers. In a subsequent Section \ref{subsec:sources}, we examine which sources are not cited with URL. How non-peer reviewed sources are used in papers is discussed in \ref{subsec:ontheuse}. In Section \ref{subsec:positional} we analyze which references tend to be cited in different parts of papers. 

\subsection{Independent researchers as data sources}
\label{subsec:independent}

One of the goals of this research was to identify the use of private blogs likely to be associated with well-informed individuals or academics, with the aim of determining whether there are dominant authors there. However, our results did not identify any such sources, apart from "kerbsonsecurity.com". Although our methodology and data collection effort should have identified these sources, as authors we wanted to re-examine why private blogs were missing from our final list. 

To identify private blogs, we took a list of all URLs in our dataset and searched them again, this time looking for the keyword "blog" within the string of each URL. This time we did not check whether the blog was about computer security, and decided to analyze all 12560 URLs found in conference papers. We identified 240 domains that contained the string "blog", but we could not identify  recurring sources. To be more precise, 173 URLs containing the string "blog" were cited only once, and 37 were cited twice, implying that private blogs do not provide consistent information on computer security, at least in 2020.

Our second guess was that there might be authors posting via online social networks such as Twitter or Medium. Searching for the keyword "twitter.com" in the URL strings, we obtained only 11 articles with links to different authors. Searching for the keyword "medium.com" yielded 28 citations in articles, but again with no dominant authors. Similarly, "youtube.com" is cited in 15 papers, also without dominant authors.

\subsection{Sources without URL in citation}
\label{subsec:sources}

Sources such as Blackhat conferences are often not cited with a URL, as described in Section \ref{subsec:identification}. For identification of this type of sources we searched URL strings against a set keywords of potential sources. After using a few sources for top hacking conferences and searching, we identified only two cited conferences: the Blackhat series with 54 citations and the DEF CON with 7 citations. At first glance, the Blackhat conference series is the best positioned source, but when analyzing the content of these citations, it becomes clear that these citations mostly focus on specific exploits and techniques, rarely on incidents as such. Since this conference is not peer-reviewed, we decided to include it in our list because it is relevant to our research topic. We could not identify any other references to conferences that were non-peer reviewed. 

A similar observation regarding citation by URL can be applied to the case of the "SANS Institute". Publications originating from this institution are sometimes cited by URL, but sometimes by document name or authors, making automatic identification difficult. We could not identify any other relevant data source which is used in the papers but without mentioning the URL in the citation.

\subsection{On the use of non-peer reviewed sources}
\label{subsec:ontheuse}

In this paper, we analyzed the top conferences that took place in 2020, so we cannot provide information about the longevity of these sources. We hypothesize that these non-peer reviewed sources change over time, and yesterday's sources do not have to provide the same quality of information. For example, we observed that the cited source "bits-please.blogspot.com" was only active from 2014 to 2016 and is still cited in 2020, but with no new information since then. In addition, all citations are concentrated on a single blog post, so this source is not a constant source of new information, but rather a one-time source. 

Newspapers with general content were found to be relevant sources of information with a total of 82 citations. These newspapers are reputable sources of trustworthy news, they are more frequently cited by authors compared to specialized news sites. These unspecialized news sources do not necessarily cover all computer security topics. However, they do cover important events and can therefore be considered as a source of valid arguments about a particular problem or incident. In this paper, we did not focus on the journalists themselves as a source of data, but only the news sites as such were analyzed.

\subsection{Positional citation of non-peer reviewed references.}
\label{subsec:positional}

Analyzing the references used, we can see that the Blackhat conference series is the main source in all article parts, except the discussion part, as shown in Table \ref{table6}. The introduction part of the articles is mainly composed of specialized or industrial sources, with 56\% of all citations coming from the top 10 sources. In the development section of the articles, the percentage is similar, with 56.7\% of citations coming from the top 10 sources, again mainly from specialized news sites and industrial sources. In the main research, we observed similar numbers, with 68\% of the citations coming from the top 10 sources. Interestingly, the list of sources for the main research consists mainly of specialized and general news sites with only one industry source in the top 10. The original hypothesis of the authors was that industry sources are mostly used in this part of the articles as they provide technical details, but our data shows otherwise. Since the discussion section of the articles has the fewest citations overall, the results of this part of the study are less reliable and we were only able to identify the top 7 sources, since all other sources are cited only once in this section. These top 7 sources cover 66.7\% of all citations and the list consists of both news and security company sources. 

One of the limiting factors in future exploration of older content is the sole access to conference proceedings. We observed that some conference websites are no longer maintained. Access to old proceedings is limited due to inactive URLs and searching for individual articles is time consuming.

\begin{table*}
\centering
\caption{Top 10 sources in regard of positional citing in papers}
\begin{tabular}{|c|c|c||c|c||c|c||c|c|} 
\hline
\textbf{No}. & \begin{tabular}[c]{@{}c@{}} \textbf{Introduction or} \\ \textbf{Related Work} \end{tabular}& \textbf{Cit.} & \textbf{Development}     & \textbf{Cit.} & \textbf{Main research}        & \textbf{Cit.} & \textbf{Discussion}        & \textbf{Cit.}       \\ 
\hline
\hline
1.  & Blackhat series              & 33   & Blackhat series  & 9   & Blackhat series      & 12   & googleprojectzero & 4          \\ 
\hline
2.  & wired.com                    & 21   & grsecurity.net   & 5   & techcrunch.com       & 6    & zdnet.com         & 3          \\ 
\hline
3.  & googleprojectzero            & 18   & forbes.com       & 5   & zdnet.com            & 5    & forbes.com        & 3          \\ 
\hline
4.  & Broadcom (Symantec)          & 18   & Fireeye          & 4   & BleepingComputer.com & 5    & theguardian.com   & 2          \\ 
\hline
5.  & zdnet.com                    & 15   & Palo Alto Networks & 4 & arstechnica.com      & 4    & security.googleblog.com   & 2          \\ 
\hline
6.  & arstechnica.com              & 11   & googleprojectzero  & 3 & theverge.com         & 3    & Microsoft (Azure, MSDN) & 2          \\ 
\hline
7.  & Kaspersky (Securelist)       & 10   & arstechnica.com  & 3   & theguardian.com      & 3    & Palo Alto Networks      & 2          \\ 
\cline{1-9}
8.  & techcrunch.com               & 8    & techcrunch.com   & 3    & DEF CON             & 3    & \multicolumn{2}{c}{\multirow{3}{*}{}}  \\ 
\cline{1-7}
9.  & theguardian.com              & 7    & theverge.com     & 3    & security.googleblog.com & 3 & \multicolumn{2}{c}{}           \\ 
\cline{1-7}
10. & security.googleblog.com      & 7    & Cisco (Talos)    & 3    & bbc.com              & 3    & \multicolumn{2}{c}{}    \\
\cline{1-7}
\end{tabular}
\label{table6}
\end{table*}

\section{Recommendations for researchers}
\label{sec:recommendations}

Our research analyzed a period of one year and has analyzed top nine computer security conferences. During our research, we found that non-peer reviewed sources are not evenly present among the conferences. When we look at the distribution of usage among conferences shown in \ref{table2}., it is clear that some conferences do not use non-peer reviewed sources: SIGCOMM, CRYPTO, FC, and ICC. This could be due to editorial practices or due to content of these conferences. The use of non-peer reviewed sources is most common in the following conferences: S\&P, Security, CCS, NDSS, and ACSAC.

The cumulative results presented in Table \ref{table6}. show that conference papers authors rely on a relatively small number of non-peer reviewed sources. In the top ten list we have:

\begin{itemize}
    \item one conference covering technical topics (Blackhat),
    \item two security companies publications (Google Project Zero and Broadcom),
    \item four specialized news sites (ZDnet, Wired, Ars Technica, TechCrunch), 
    \item three news websites (The Guardian, The Verge, Forbes).
\end{itemize}

In addition, we provide researchers with information on how these references were used. Not all sources contain data relevant to the research itself. Overall, non-peer reviewed sources are most often, 61\% of the time, used by researchers as motivation for the work (see Table \ref{table5} for full results). Some sources are more frequently used for motivation, but others contain data that is more frequently used in the main part of the research, as shown in the overview in Table \ref{table6}. These sources are accepted by reviewers in top computer security conferences and as such should be followed by computer security researchers as they publish trustworthy information. We recommend researchers to follow these sources as they can provide important information for future research.

\section{Conclusion and Future Work}
\label{sec:conclusion}

In this paper, we have developed a methodology for identifying and analyzing the use of non-peer reviewed sources in computer security. To the best of our knowledge, this has not been done before. We analyzed the top computer security conferences of 2020, as conferences tend to have a shorter review time compared to journals. Our results show that different conferences tend to show different levels of reliance on non-peer reviewed sources. Of the nine conferences analyzed, five conferences contained references to 97\% of analyzed citations, while we did not identify significant use of non-peer reviewed sources in four conferences. 
We identified a total of 37 different non-peer reviewed sources with more than 5 citations in the observed conferences. Non-peer reviewed sources are used to motivate the work in 61\% of the cases, followed by the development section with 17\% and the main research with 16\%, with the remaining 6\% falling into the discussion section. Research suggests that different sources tend to be dominant in different sections of articles, presumably because they do not provide the same type of information. 

Most of the non-peer reviewed sources are specialized websites, security company publications, and organizational blogs. These sources are trusted enough to be allowed in top conferences and as such can be seen as sources of trustworthy information on computer security and should therefore be considered by researchers. 

During our research, we identified only one relevant "well informed" individual or small organization who publishes security news. Both private and organizational blogs were cited a total of 421 times, but they are scattered across 240 different sources.

In this paper we did not conduct a longitudinal study, and it would be interesting to determine both: how these data correlate over time and whether there is a correlation between non-peer reviewed data sources used in journals and conferences. In addition, at this point we do not know how this data compares with differently rated conferences and journal articles. It could be that lower rated conferences are later cited in higher rated conferences, thus used as information filters.

\bibliographystyle{ACM-Reference-Format}
\bibliography{bibliography}

\end{document}